\documentclass[aip,cha,preprint,onecolumn,superscriptadress,english]{revtex4-1}

\usepackage{amsmath}
\usepackage{amssymb}
\usepackage{amsthm}
\usepackage{wasysym}
\usepackage[latin1]{inputenc}
\usepackage[T1]{fontenc}
\usepackage{ae,aecompl}

    \usepackage[dvips]{graphicx}
    \DeclareGraphicsExtensions{.eps}

\newcommand{\be}{\begin{equation}}
\newcommand{\ee}{\end{equation}}
\newcommand{\bea}{\begin{eqnarray}}
\newcommand{\eea}{\end{eqnarray}}
\newcommand{\nn}{\nonumber}

\begin{document}
\bibliographystyle{apsrev}
\title{Impact of network topology on synchrony of oscillatory power grids}

\author{Martin Rohden}
\affiliation{Network Dynamics, Max Planck Institute for Dynamics and Self-Organization (MPIDS),
37077 G\"ottingen, Germany}

\author{Andreas Sorge}
\affiliation{Network Dynamics, Max Planck Institute for Dynamics and Self-Organization (MPIDS),
37077 G\"ottingen, Germany}

\author{Dirk Witthaut}
\affiliation{Network Dynamics, Max Planck Institute for Dynamics and Self-Organization (MPIDS),
37077 G\"ottingen, Germany}

\author{Marc Timme}
\affiliation{Network Dynamics, Max Planck Institute for Dynamics and Self-Organization (MPIDS),
37077 G\"ottingen, Germany}
\affiliation{Faculty of Physics, Georg August University G\"ottingen, Germany}

\date{\today}

\begin{abstract}

Replacing conventional power sources by renewable sources in current power
grids drastically alters their structure and functionality. The resulting
grid will be far more decentralized with an distinctly different topology.
Here we analyze the impact of grid topologies on spontaneous
synchronization, considering regular, random and small-world topologies
and focusing on the influence of decentralization. We employ the
\textit{classic model} of power grids modeling consumers and sources as
second order oscillators. We first analyze the global dynamics of the
simplest non-trivial (two-node) network that exhibit a synchronous (normal
operation) state, a limit cycle (power outage) and coexistence of both.
Second, we estimate stability thresholds for the collective dynamics of
small network motifs, in particular star-like networks and regular grid
motifs. For larger networks we numerically investigate decentralization
scenarios finding that decentralization itself may support power grids in 
reaching stable synchrony for lower lines transmission capacities.
Decentralization may thus be beneficial for power grids, regardless of
their special resulting topology. Regular grids exhibit a specific
transition behavior not found for random or small-world grids.
\end{abstract}

\pacs{05.45.Xt,84.70.+p,89.75.-k}

\maketitle


\begin{quotation}
The availability of electric energy fundamentally underlies all aspects of life; thus its
reliable distribution is indispensable. The drastic change from our traditional energy
system based on fossil fuels to one based dominantly on renewable sources provides an
extraordinary challenge for the robust operation of future power grids \cite{Butl07}. Renewable sources are intrinsically smaller and more decentralized,
thus yielding connection topologies strongly distinct from those of today. How network topologies impact the collective dynamics and in particular
the stability of standard grid operation, is still not well understood. In this article, we systematically study how decentralization may
influence collective grid dynamics in model oscillatory networks. We first study small network motifs that serve as model system for the larger networks,
that are analyzed in this article. We find that, independent of global topological features, decentralized grids are consistenly able to reach
their stable state for lower transmission line capacities. At least regarding pure topological issues, decentralizing grids may thus be beneficial for
operating oscillatory power grids, largely independent of both the original and the resulting grid.
\end{quotation}

\section{Introduction}
The compositions of current power grids undergo radical changes. As of now, power grids are still dominated by big conventional power
plants based on fossil fuel or nuclear power exhibiting a large power output. Essentially, their effective topology is locally star-like with transmission lines going from large plants to regional
consumers. As more and more renewable power sources contribute, this is about to change and topologies will become more decentralized and more
recurrent. The topologies of current grids largely vary, with large differences, e.g. between grids on islands such as Britain and those in
continental Europe, or between areas of different population densities. In addition, renewable sources will strongly modify these structures in a yet
unknown way. The synchronization dynamics of many power grids with a special topology are well analyzed \cite{Motter13}, such as the British power grid \cite{Witt12}
or the European power transmission network \cite{Loza12}. The general impact of grid topologies on collective dynamics is not systematically understood, in particular with respect to
decentralization.

Here, we study collective dynamics of oscillatory power grid models with a special focus on how a wide range of topologies, regular, small-world and random,
influence stability of synchronous (phase-locked) solutions. We analyze the onset of phase-locking between power generators and consumers as well as the local and global stability of
the stable state. In particular, we address the question of how phase-locking is affected in different topologies if large power plants are replaced
by small decentralized power sources. For our simulations, we model the dynamics of the power grid as a network of coupled second-order oscillators, which are
derived from basic equations of synchronous machines \cite{Fila08}. This model bridges the gap between large-scale static network models \cite{Mott02,Scha06,Simo08,Heid08} on
the one hand and detailed component-level models of smaller network \cite{Eurostag} on the other. It thus admits systematic access to emergent
dynamical phenomena in large power grids. 

The article is organized as follows. We present a dynamical model for power grids in Sec.~\ref{sec-model}. The basic dynamic properties, including
stable synchronization, power outage and coexistence of these two states, are discussed in Sec.~\ref{sec-smallnets} for elementary networks. These
studies reveal the mechanism of self-organized synchronization in a power grid and help understanding the dynamics also for more complex
networks. In Sec.~\ref{sec-largenets} we present a detailed analysis of large power grids of different topologies. We investigate the onset of
phase-locking and analyze the stability of the phase-locked state against perturbations, with an emphasis on how the dynamics depends on the
decentralization of the power generators. Stability aspects of decentralizing power networks has been briefly reported before for the British
transmission grid \cite{Witt12}.

\section{Coupled oscillator model for power grids}
\label{sec-model}

We consider an oscillator model where each element is one of two types of elements, generator or consumer \cite{Fila08,Prab94}. Every element $i$ is described by the
same equation of motion with a parameter $P_i$ giving the generated $(P_i>0)$ or consumed $(P_i<0)$ power. The state of each element is determined
by its phase angle $\phi_i (t)$ and velocity $\dot\phi_i(t)$. During
the regular operation, generators as well as consumers within the grid run with the same frequency $\Omega =2\pi\times 50\text{Hz}$ or
$\Omega =2\pi\times 60\text{Hz}$. The phase of each element $i$ is then written as
\begin{equation}
  \phi_i(t)=\Omega t + \theta_i(t) \label{eqn:phase},
\end{equation}
where $\theta_i$ denotes the phase difference to the set value $\Omega t$.\\
The equation of motion for all $\theta_i$ can now be obtainend from the energy conservation law, that is the generated or consumed energy
$P^{\text{source}}_i$ of each single element must equal the energy sum given or taken from the grid plus the accumulated and dissipated energy of
this element. The dissipation power of each element is $P^{\text{diss}}_i=\kappa_i(\dot\phi_i)^2$, the accumulated power\
$P^\text{acc}_i=\frac{1}{2}I_i\frac{d}{dt}(\dot\phi_i)^2$ and the transitional power between two elements is
$P^{\text{trans}}_{ij}=-P^{\text{max}}_{ij}\sin(\phi_j-\phi_i)$. Therefore $P^{\text{source}}_i$
is the sum of these:
\begin{equation}
P^{\text{source}}_i= P^{\text{diss}}_i+ P^{\text{acc}}_i+ P^{\text{trans}}_{ij}\label{eqn:grund}.
\end{equation}
An energy flow between two elements is only possible if there is a phase difference between these two. Inserting equation (\ref{eqn:phase}) and
assuming only slow phase changes compared to the frequency $\Omega$ $(|\dot\theta_i| \ll\Omega)$. The dynamics of the $i$th machine is given by:
\begin{equation}
I_i\Omega\ddot\theta_i=P^{\text{source}}_i
     -\kappa_i\Omega^2-2\kappa_i\Omega\dot\theta_i+\sum_j P^{\text{max}}_{ij}\sin(\theta_j-\theta_i).
\end{equation}
Note that in this equation only the phase differences $\theta_i$ to the fixed phase $\Omega t$ appear. This shows that only the phase difference
between the elements of the grid matters. The elements $K_{ij}=\frac{P^{\text{max}}_{ij}}{I_i\Omega}$ constitute the connection matrix of the
entire grid, therefore it decodes wether or not there is a transmission line between two elements ($i$ and $j$). With
$P_i=\frac{P^{\text{source}}_i-\kappa_i \Omega^2}{I_i\Omega}$ and $\alpha_i=\frac{2 \kappa_i}{I_i}$ this leads to the following equation of motion:
\begin{equation}
\frac{d^2\theta_i}{dt^2}=P_i-\alpha_i\frac{d\theta_i}{dt}+\sum_j K_{ij}\sin(\theta_j-\theta_i).
\end{equation}
The equation can now be rescaled with $s=\alpha t$ and new variables $\tilde P=P/ \alpha^2$ and $\tilde K=K/ \alpha^2$. This leads to:
\begin{equation}
\frac{d^2\theta_i}{ds^2}=\tilde P_i-\frac{d\theta_i}{ds}+\sum_j \tilde K_{ij}\sin(\theta_j-\theta_i).
\end{equation}
In the stable state both derivatives $\frac{d\theta_i}{dt}$ and $\frac{d^2\theta_i}{dt^2}$ are zero, such that
\begin{equation}
 0=P_i + \sum_j K_{ij}\sin(\theta_j-\theta_i)
\end{equation}
holds for each element in the stable state. For the sum over all equations, one for each element $i$, the following holds
\begin{equation}
 \sum_i P_i=\sum_{i<j}K_{ij}\sin(\theta_j-\theta_i)+\sum_{i>j}K_{ij}\sin(\theta_j-\theta_i)=0, \label{eqn:sum}
\end{equation}
because $K_{ij}=K_{ji}$ and the sin-function is antisymmetric. This means it is a necessary condition that the sum of the generated power
$(P_i>0)$ equals the sum of the consumed power $(P_i<0)$ in the stable state.\\
For our simulations we consider large centralized power plants generating $P^{\text{source}}_i = 100 \, {\rm MW}$ each. A synchronous generator of this
size would have a moment of inertia of the order of $I_i=10^4 \, {\rm kg \, m}^2$. The mechanically dissipated power $\kappa_i\Omega^2$ usually is a small
fraction of $P^{\text{source}}$ only. However, in a realistic power grid there are additional sources of dissipation, especially ohmic losses and
because of damper windings \cite{Macho}, which are not taken into account directly in the coupled oscillator model. Therefore we set
$\alpha_i=0.1 s^{-1}$ and $P_i = 10 s^{-2}$ \cite{Fila08} for large power plants. For
a typical consumer we assume $P_i = -1 s^{-2}$, corresponding to a small city. For a renewable power plant we assume $P_i=2.5 s^{-2}$. A major overhead power line can
have a transmission capacity of up to $P^{\text{max}}_{ij} = 700 \, {\rm MW}$. A power line connecting a small city usually has a smaller transmission
capacity, such that $K_{ij} \le 10^2 s^{-2}$ is realistic. We take $\Omega = 2\pi\times 50\text{Hz}$.

\section{Dynamics of elementary networks}
\label{sec-smallnets}

\subsection{Dynamics of one generator coupled with one consumer}
\label{sec-two-elements}

\begin{figure}[t]
\centering
\includegraphics[width=10cm,  angle=0]{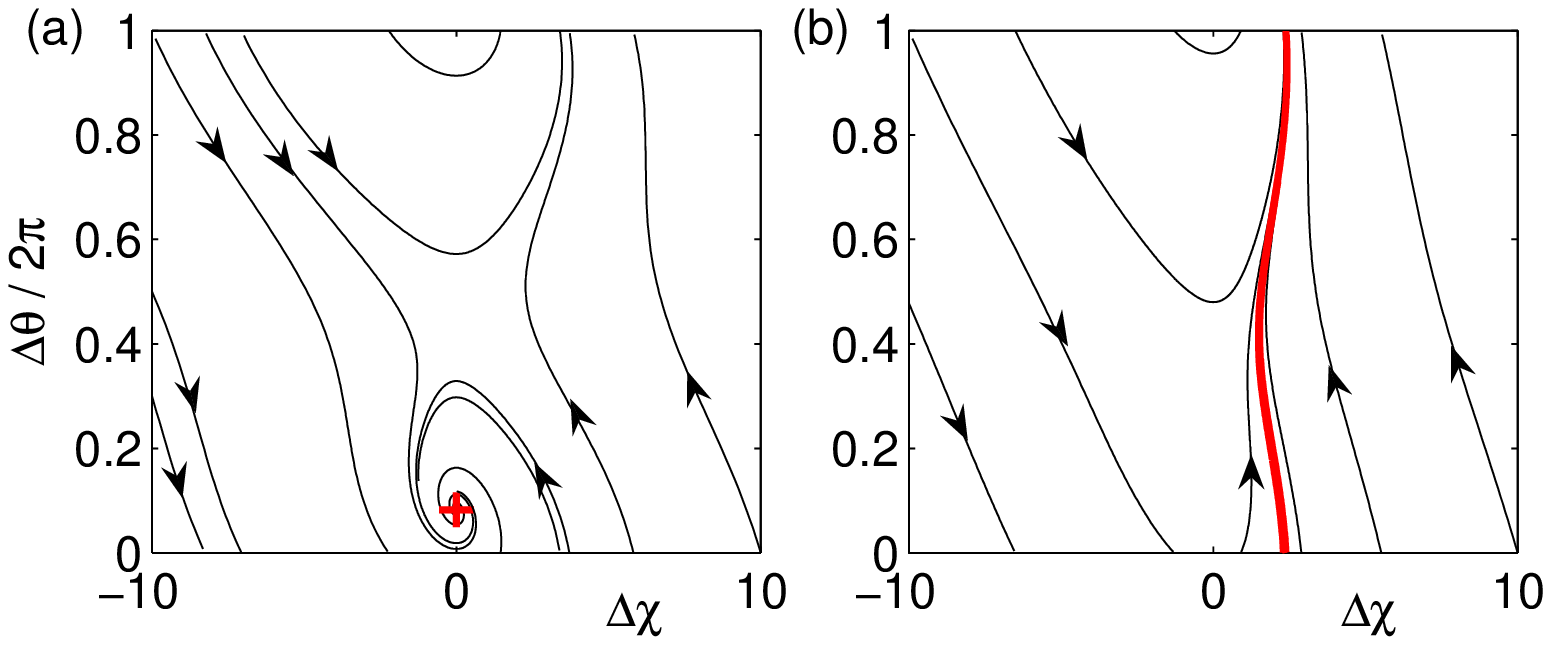}
\includegraphics[width=10cm,  angle=0]{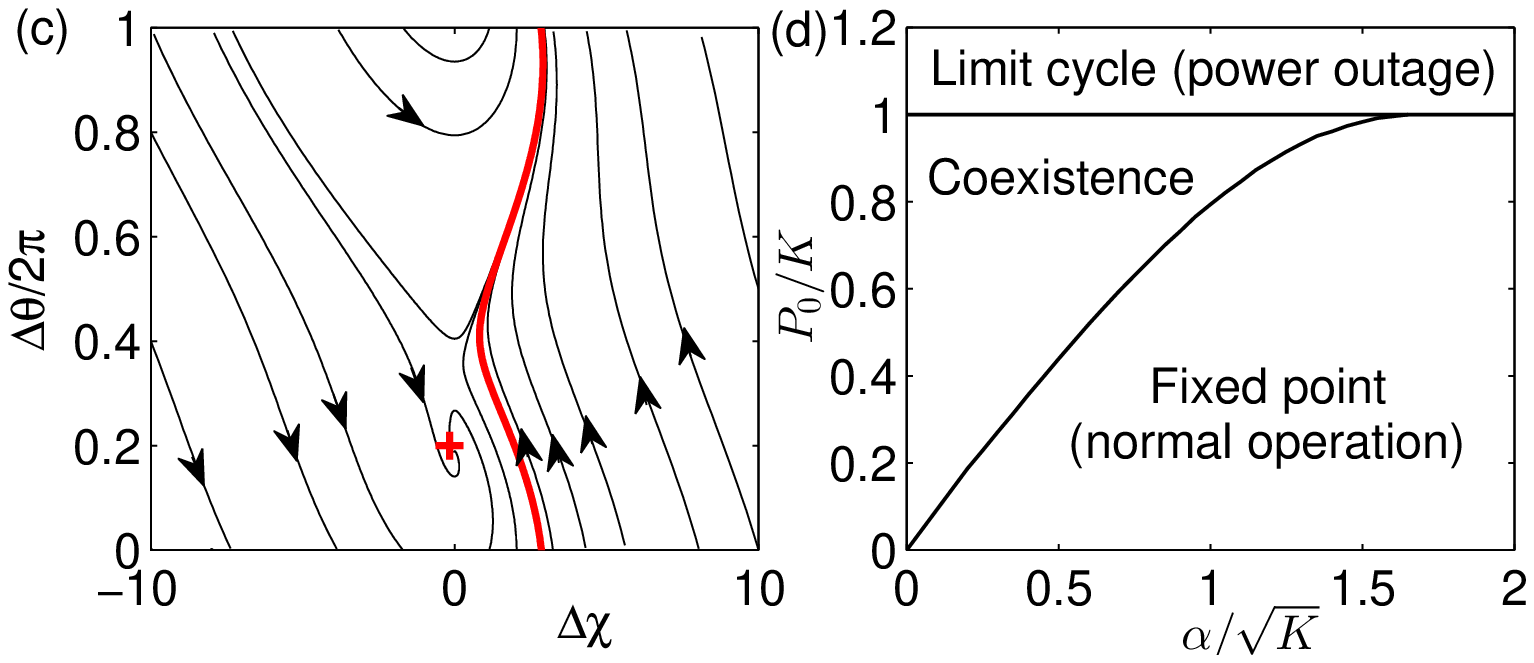}
\caption{\label{fig:globalphase}
Dynamics of an elementary network with one generator and one consumer for $\alpha=0.1 s^{-1}$.\\
(a) Globally stable phase locking for $P_0 = 1 s^{-2}$ and $K = 2 s^{-2}$\\
(b) Globally unstable phase locking (limit cycle) for $P_0 = 1 s^{-2}$ and $K = 0.5 s^{-2}$\\
(c) Coexistence of phase locking (normal operation) and limit cycle (power outage) for $P_0 = 1 s^{-2}$ and $K = 1.1 s^{-2}$\\
(d) Stability phase diagram in parameter space.
}
\end{figure}

We first analyze the simplest non trivial grid, a two-element system consisting of one generator and one consumer. This system is analytically
solvable and reveals some general aspects also present in more complex systems. This system can only reach equilibrium if equation (\ref{eqn:sum})
is satisfied, such that $-P_1 = P_2$ must hold. With $\Delta P=P_2-P_1$ the equation of motion for this system can be simplified in such a way, that only the phase difference
$\Delta\theta=\theta_2-\theta_1$ and the difference velocity $\Delta\chi:=\Delta\dot\theta$ between the oscillators is decisive:
\begin{eqnarray}
\Delta\dot\chi&=&\Delta P-\alpha\Delta\chi-2K\sin\Delta\theta \label{eqn:eom}\nonumber \\
\Delta\dot\theta&=&\Delta\chi.
\end{eqnarray}
Figure \ref{fig:globalphase} shows different scenarios for the two-element system. For $2 K \geq \Delta P$ two fixed points come into being
\ref{fig:globalphase}(a), whose local stability is analyzed in detail below. The system is globally stable as is shown in the bottom area of
Fig.~\ref{fig:globalphase} (d). For $2K < \Delta P$ the load exceeds the capacity of the link. No stable operation is possible and all trajectories
converge to a limit cycle as shown in fig \ref{fig:globalphase} (b) and in the upper area of \ref{fig:globalphase} (d). In the remaining region
of parameter space, the fixed point and the limit cycle coexist such that the dynamics depend crucially on the initial conditions as shown in
fig \ref{fig:globalphase} (c) (cf.~\cite{Risk96}). Most major power grids are operating close to the edge of stability, i.e. in the region of
coexistence, at least during periods of high loads. Therefore the dynamics depends crucially on the initial conditions and static power grid models
are insufficient.

Let us now analyze the fixed points of the equations of motion (\ref{eqn:eom}) in more detail. In terms of the phase difference $\Delta\theta$,
they are given by:
\begin{eqnarray}
T_1&:&\quad \begin{pmatrix}\Delta\chi^*\\\Delta\theta^*\end{pmatrix}=\begin{pmatrix}0\\ \arcsin\frac{\Delta P}{2K}\end{pmatrix}, \nonumber \\
T_2&:&\quad \begin{pmatrix}\Delta\chi^*\\\Delta\theta^*\end{pmatrix}=\begin{pmatrix}0\\ \pi -\arcsin\frac{\Delta P}{2K}\end{pmatrix}.
\end{eqnarray}
For $\Delta P > 2K$ no fixed point can exist as discussed above. The critical coupling strengths $K_c$ is therefore $\Delta P/2$. Otherwise fixed points exist and the system can reach a stationary state. For
$\Delta P=2K$ only one fixed point exists, $T_1=T_2$, at $\left(\Delta\chi^*,\Delta\theta^*\right)=(0,\pi/2)$. It is neutraly stable.\\
We have two fixed points for $2K>\Delta P$. The local stability of these fixed points is
determined by the eigenvalues of the Jacobian of the dynamical system (\ref{eqn:eom}), which are given by
\begin{equation}
\lambda_{\pm}^{(1)}=-\frac{\alpha}{2}\pm\sqrt{\left(\frac{\alpha}{2}\right)^2-\sqrt{4K^2-\Delta P^2}}
\end{equation}
at the first fixed point $T_1$ and
\begin{equation}
\lambda_{\pm}^{(2)}=-\frac{\alpha}{2}\pm\sqrt{\left(\frac{\alpha}{2}\right)^2+\sqrt{4K^2-\Delta P^2}}
\end{equation}
at the second fixed point $T_2$, respectively.
Depending on $K$, the eigenvalues at the first fixed point are either both real and negativ or complex with negative real values. One
eigenvalue at the second fixed point is always real and positive, the other one real and negative. Thus only the first fixed point is stable and enables a stable
operation of the power grid. It has real and negative eigenvalues for $K_c < K < K_2=\sqrt{\frac{\alpha^4}{64}+\frac{\Delta P^2}{4}}$, which is only
possible for large $\alpha$, i.e. the system is overdamped. For $K \geq K_2$ it has complex eigenvalues with a negative real value $|\Re(\lambda)|\equiv \frac{\alpha}{2}$,
for which the power grid exhibits damped oscillations around the fixed point. As power grids should work with only minimal losses, which corresponds to $K \geq K_2$, this is the
practically relevant setting.\\

\subsection{Dynamics of motif networks}
\label{sec-starnet}

\begin{figure}[t]
\centering
\includegraphics[width=8cm,  angle=0]{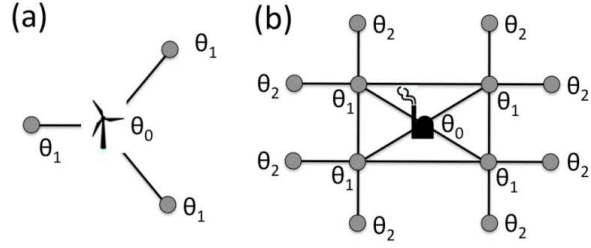}
\caption{\label{fig-starnet} Motif networks: simplified phase description}
\end{figure}

We discuss the dynamics of the two motif networks shown in Fig.~\ref{fig-starnet}. These two can be considered as building
blocks of the large-scale quasi-regular network that will be analyzed in the next section.
Fig.~\ref{fig-starnet} (a) shows a simple network, where a small renewable energy source provides the power for $N=3$ consumer units with $d=3$ connections. To analyze
the most homogeneous setting we assume that all consumers have the same phase $\theta_1$ and a power load of $-P_0$ and all transmission lines have the
same capacity $K$. The power generator has the phase $\theta_0$ and provides a power of $N P_0$. The reduced equations of
motion then read
\bea
   \ddot \theta_0 &=& N \, P_0 - \dot \theta_0 
        + d K \sin(\theta_1 - \theta_0), \nn \\
    \ddot \theta_1 &=& - P_0 -  \dot \theta_1
        + K \sin(\theta_0 - \theta_1)
\eea.
For this motif class the condition $|N|=|d|$ always holds, such that the steady state is determined by $\sin(\theta_0 - \theta_1) = P_0/K$. The condition for the existence of a steady state is thus
$K > K_c = P_0$, i.e. each transmission line must be able to transmit the power load of one consumer unit.

Fig.~\ref{fig-starnet} (b) shows a different network, where $N=12$ consumer units arranged on a squared lattice with $d_1=4$ connections between
the central power source $(\theta_0)$ and the nearest consumers ($\theta_1$) and $d_2=2$ connections between the consumers with phase $\theta_1$ and those with
$\theta_2$. Due to the symmetry of the problem we have to consider only three different phases. The reduced equations
of motion then read
\bea
   \ddot \theta_0 &=& N \, P_0 -  \dot \theta_0 
        + d_1 K \sin(\theta_1 - \theta_0), \nn \\
   \ddot \theta_1 &=& - P_0 -  \dot \theta_1
       + d_2 K \sin(\theta_2 - \theta_1)    + K \sin(\theta_0 - \theta_1), \nn \\
    \ddot \theta_2 &=& - P_0 - \dot \theta_2
        + K \sin(\theta_1 - \theta_2). 
\eea
For the steady state we thus find the relations
\bea
    \sin(\theta_0 - \theta_1) &=& (N P_0)/(d_1K) \nn \\
    \sin(\theta_1 - \theta_2) &=& P_0/K.
\eea
The coupling strengths $K$ must now be higher than the critical coupling strenghts
\begin{equation}
  K_c=\frac{NP_0}{d_1}\label{eqn:coupling}
\end{equation}
to enable a stable operation. For the example shown in Fig.~\ref{fig-starnet} (b) we now have a higher critical coupling strength $K_c = 3 P_0$
compared to the previous motif for the existence of a steady state. This is immediately clear from physical reasons, as the transmission lines
leading away from the power plant now have to serve 3 consumer units instead of just one.

\section{Dynamics of large power grids}
\label{sec-largenets}

\subsection{Network topology}

\begin{figure}[t]
\centering
\includegraphics[width=7cm, angle=0]{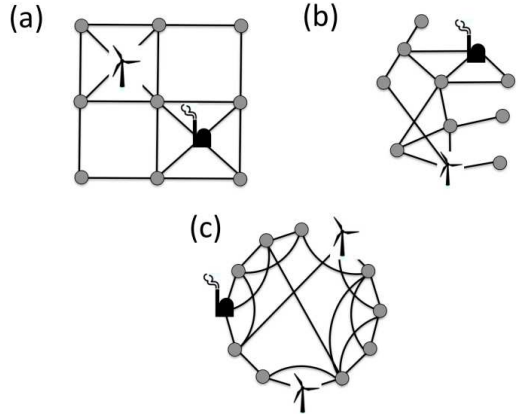}
\caption{\label{fig-network} Small size cartoons of different network topologies: (a) Quasi-regular grid, (b) random network
and (c) small-world network.}
\end{figure}

We now turn to the collective behavior of large networks of coupled generators and consumers and analyze how the dynamics and stability of a power
grid depend on the network structure. We emphasize how the stability is affected when large power plants are replaced by many small
decentralized power sources.

In the following we consider power grids of $N_C = 100$ consumers units with the same power load $-P_0$ each. In all simulations we
assume $P_0 = 1 s^{-2}$ with $\alpha=0.1 s^{-1}$ as discussed in Sec.~\ref{sec-model}. The demand of the consumers is met by
$N_P \in \{0,\ldots,10\}$ large power plants, which provide a power $P_P = 10 \, P_0$ each. The remaining power is generated by $N_R$ small
decentralized power stations, which contribute $P_R = 2.5 \, P_0$ each. Consumers and generators are connected by transmission lines with a
capacity $K$, assumed to be the same for all connections.\\ We consider three types of networks topologies, schematically
shown in Fig.~\ref{fig-network}. In a quasi-regular power grid, all consumers are placed on a squared lattice. The generators are placed randomly at
the lattice and connected to the adjacent four consumer units (cf. Fig.~\ref{fig-network} (a)). In a random network, all elements
are linked completely randomly with an average number of six connections per node (cf. Fig.~\ref{fig-network} (b)). A small world network is
obtained by a standard rewiring algorithm \cite{Watt98} as follows. Starting from ring network, where every element is connected to its four nearest
neighbors, the connections are randomly rewired with a probability of $0.1$ (cf. Fig.~\ref{fig-network} (c)).

\subsection{The synchronization transition}

\begin{figure}[t]
\centering
\includegraphics[width=10cm,  angle=0]{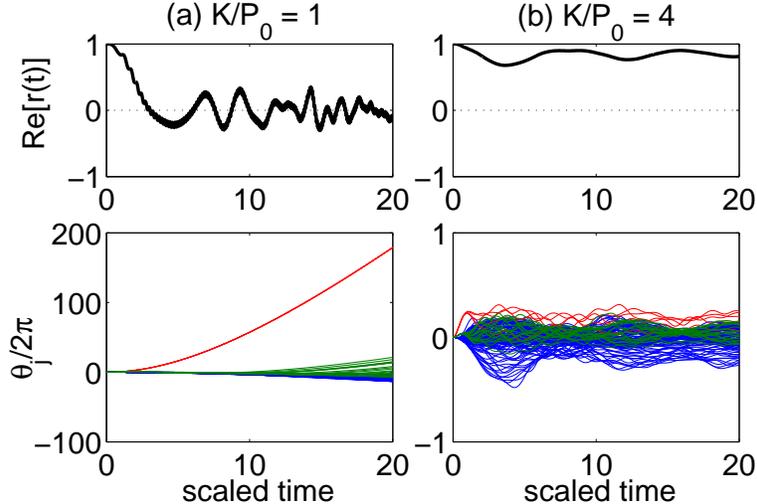}
\caption{\label{fig-sync} Synchronization dynamics of a quasi-regular power grid. (a) For a weak coupling the phases $\theta_j(t)$ of the small renewable
decentralized generators (green lines) synchronize with the consumers (blue lines), but not the phases of the large power plants (red lines).
Thus the order parameter $r(t)$ fluctuates around a zero mean. (b) Global phase-locking of all generators and  consumers is achieved for a large
coupling strength, such that the order parameters $r(t)$ is almost one.}
\end{figure}

We analyze the requirements for the onset of phase locking between generators and consumers, in particular the minimal coupling
strength $K_c$. An example for the synchronization transition is shown in Fig.~\ref{fig-sync}, where the dynamics of the phases $\theta_i(t)$ is
shown for two different values of the coupling strength $K$. Without coupling, $K=0$, all elements of the grid oscillate with their natural frequency. For small values of $K$, synchronization sets in between the renewable generators and the consumers whose frequency difference is
rather small (cf. Fig.~\ref{fig-sync} (a)). Only if the coupling is further increased (Fig.~\ref{fig-sync} (b)), all generators synchronize so
that a stable operation of the power grid is possible.

\begin{figure}[t]
\centering
\includegraphics[width=10cm,  angle=0]{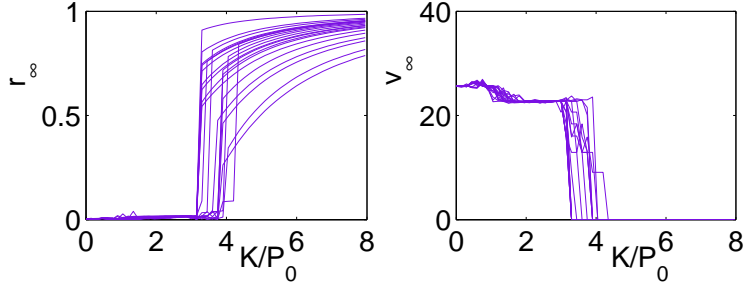}
\caption{\label{fig-order-fluct} 
The synchronization transition as a function of the coupling strength $K$:
The order parameter $r_{\infty}$ (left-hand side) and the phase velocity 
$v_{\infty}$ (right-hand side) in the long time limit.
The dynamics has been simulated for 20 different realizations of a quasi-regular
network consisting of 100 consumers, $N_P=6$ large power pants and $N_R=16$ small power 
generators.
}
\end{figure}

The phase coherence of the oscillators is quantified by the order parameter  \cite{Stro00}
\be
  r(t)  =  \frac{1}{N} \sum \nolimits_j e^{i \theta_i(t)},
\ee
which is also plotted in Fig.~\ref{fig-sync}. For a synchronous operation, the real part of the order parameters is almost one, while it fluctuates around
zero otherwise. In the long time limit, the system will either relax to a steady synchronous state or to a limit cycle where the generators and
consumers are decoupled and $r(t)$ oscillates around zero. In order to quantify synchronization in the long time limit we thus define the
averaged order parameter
\be
  r_{\infty} :=  \lim_{t_1 \rightarrow \infty}  \lim_{t_2 \rightarrow \infty} 
   \frac{1}{t_2} \int^{t_1+t_2}_{t_1} r(t) \, dt.
\ee
In numerical simulations the integration time $t_2$ must be finite, but large compared to the oscillation period if the system
converges to a limit cycle. Furthermore we consider the averaged squared phase velocity
\be
  v^2(t) =  \frac{1}{N} \sum \nolimits_j \dot \theta_j(t)^2,
\ee
and its limiting value
\begin{equation}
 v_\infty^2 := \lim_{t_1 \rightarrow \infty}  \lim_{t_2 \rightarrow \infty}\frac{1}{t_2}\int_{t_1}^{t_1+t_2}v^2(t)dt 
\end{equation}
as a measure of whether the grid relaxes to a stationary state.
These two quantities are plotted in Fig.~\ref{fig-order-fluct} as a function of the coupling
strength $K/P_0$ for 20 realizations of a quasi-regular network with 100 consumers and
40 \% renewable energy sources. The onset of synchronization is clearly visible: If the
coupling is smaller than a critical value $K_c$ no steady synchronized state exists 
and $r_{\infty} = 0$ by definition. Increasing $K$ above $K_c$ leads to the
onset of phase locking such that $r_{\infty}$ jumps to a non-zero value.
The critical value of the coupling strength is found to lie in the range 
$K_c/P_0 \approx 3.1 - 4.2$, depending on the random realization of
the network topology.

\begin{figure}[t]
\centering
\includegraphics[width=10cm,  angle=0]{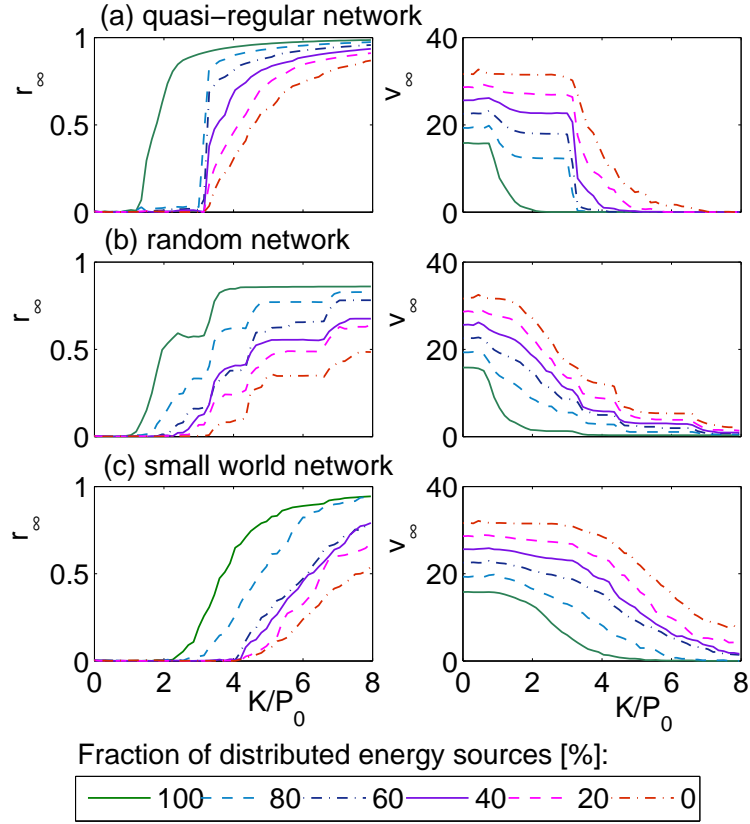}
\caption{\label{fig-order} 
The synchronization transition for different fractions of decentralized energy 
sources $1-N_P/10$ feeding the grid and for different network topologies: 
(a) Quasi-regular grid, (b) random network and (c) small-world network.
The order parameter $r_{\infty}$  and the phase velocity
$v_{\infty}$ (cf. Fig.~\ref{fig-order-fluct}) have been averaged over
100 realizations for each network structure and each fraction of decentralized sources.}
\end{figure}

The synchronization transition is quantitatively analyzed in Fig.~\ref{fig-order}. We plotted $r_{\infty}$ and
$v_{\infty}$ for three different network topologies averaged over 100 random realizations for each amount of decentralized energy sources for every
topology. The synchronization transition strongly depends on the structure of the network, and in particular the amount of power provided by small
decentralized energy sources. Each line in Fig.~\ref{fig-sync} corresponds to a different fraction of decentralized energy $1 - N_P/10$, where $N_P$ is
the number of large conventional power plants feeding the grid. Most interestingly, the introduction of small decentralized power sources
(i.e. the reduction of $N_P$) promotes the onset of  synchronization. This phenomenon is most obvious for the random and the small-worlds
structures.

Let us first analyze the quasi-regular grid in the limiting cases $N_P = 10$ (only large power plants) and $N_P=0$ (only small decentralized power
stations) in detail. The existence of a synchronized steady state requires that the transmission lines leading away from a generator have enough
capacity to transfer the complete power, i.e. $10 \, P_0$ for a large power plant and $2.5 \, P_0$ for a small power station.
In a quasi-regular grid every generator is connected with exactly four transmission lines, which leads to the following estimate for the critical
coupling strength (cf. equation \ref{eqn:coupling}):
\bea
  K_c &=& 10P_0/4 \,  \qquad \mbox{for} \, N_P = 10, \nn \\
  K_c &=& 2.5P_0/4 \, \qquad \;\; \mbox{for} \, N_P = 0.
  \label{eqn-kc-est}
\eea
These values only hold for a completely homogeneous distribution of the power load and thus rather present a lower bound for $K_c$ in a
realistic network. Indeed, the numerical results shown in Fig.~\ref{fig-order} (a) yield a critical coupling strength of
$K_c \approx 3.2 \times P_0$ and $K_c \approx 1 \times P_0$, respectively.

For networks with a mixed structure of power generators ($N_P \in \{ 1,\ldots,9\}$) we observe that the synchronization transition is determined by the
large power plants, i.e. the critical coupling is always given by $K_c \approx 3.2 \times P_0$  as long as $N_P \neq 0$. However, the transition is
now extremely sharp -- the order parameter does not increase smoothly but rather jumps to a high value. This results from the fact that all small
power stations are already strongly synchronized with the consumers for smaller values of $K$ and only the few large power plants are missing. When
they finally fall in as the coupling strength exceeds $K_c$, the order parameter $r$ immediately jumps to a large value.

The sharp transition at $K_c$ is a characteristic of the quasi-regular grid. For a random and a small-world network different classes of power
generators exist, which are connected with different numbers of transmission lines. These different classes get synchronized to the consumers
one after another as $K$ is increased, starting with the class with the highest amount of transmission lines to the one with fewest. Therefore we
observe a smooth increase of the order parameter $r$.

\subsection{Local stability and synchronization time}

\begin{figure}[t]
\centering
\includegraphics[width=10cm,  angle=0]{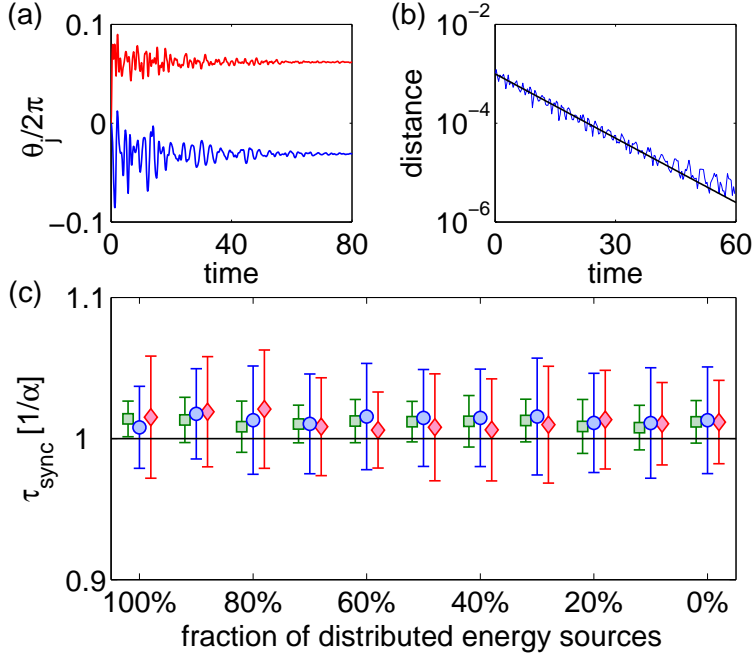}
\caption{\label{fig-relax}
Relaxation to the synchronized steady state:
(a) Illustration of the relaxation process $(K/P_0=10$ and $N_p = 10)$. We have plotted the dynamics of the phases $\theta_j$ only for one
generator (red) and one consumer (blue) for the sake of clarity.
(b) Exponential decrease of the distance to the steady state (blue line) and a fit according to $d(t) \sim e^{-t/\tau_{\rm sync}}$ (black line).
(c) The synchronisation time $\tau_{\rm sync}$  as a function of
the fraction of decentralized energy sources $1-N_P/10$
for a regular ($\circ$), a random ($\square$) and a small-world grid ($\diamond$). Cases where the system does not relax have been discarded.
}
\end{figure}

A sufficiently large coupling of the nodes leads to synchronization of all nodes of a power grid as shown in the preceding section. Starting from
an arbitrary state in the basin of attraction, the network relaxes to the stable synchronized state with a time scale $\tau_{\rm sync}$. For
instance, Fig.~\ref{fig-relax} (a) shows the damped oscillations of the phase $\theta_j(t)$ of a power plant and a consumer in a quasi-regular grid
with $K=10$ and $N_P = 10$. In order to quantify the relaxation, we calculate the distance to the steady state
\be
  d(t) =\left( \sum_{i=1}^N d_1^2(\theta_i(t),\theta_{i,\rm st})
                  + d_2^2(\dot \theta_i(t),\dot \theta_{i,\rm st})\right)^{\frac{1}{2}},
     \label{eqn-dist-tot}
\ee
where the subscript 'st' denotes the steady state values. For the phase velocities $d_2$ denotes the common Euclidian distance
$d_2^2\left(\dot\alpha,\dot\beta\right)=|\dot\alpha-\dot\beta |^2$, while the circular
distance of the phases is defined as
\be
   d_1(\alpha,\beta) = 1 - \cos(\alpha - \beta).
\ee
The distance $d(t)$ decreases exponentially during the relaxation to the steady state as shown in  Fig.~\ref{fig-relax} (b). The
black line in the figure shows a fit with the function $d(t) = d_0 \exp(-t/\tau_{\rm sync})$. Thus synchronization time $\tau_{\rm sync}$
measures the local stability of the stable fixed point, being the inverse of the stability exponent $\lambda$
(cf. the discussion in Sec.~\ref{sec-two-elements}).

Fig.~\ref{fig-relax} (c) shows how the synchronization time depends on the structure of the network and the mixture of power generators. For
several paradigmatic systems of oscillators, it was recently shown that the time scale of the relaxation process depends crucially on the network
structure \cite{Grab10}. Here, however, we have a network of {\it damped} second order oscillators. Therefore the
relaxation is almost exclusively given by the inverse damping constant $\alpha^{-1}$. Indeed we find $\tau_{\rm sync} \gtrsim \alpha^{-1}$. For
an elementary grid with two nodes only, this was shown rigorously in Sec.~\ref{sec-two-elements}. As soon as the coupling strength exceeds a critical
value $K>K_2$, the real part of the stability exponent is given by $\alpha$, independent of the other system parameters. A different value is found
only for intermediate values of the coupling strength $K_c < K < K_2$. Generally, this remains true also for a complex network of many
consumers and generators as shown in Fig.~\ref{fig-relax} (c). For the given parameter values we observe neither a systematic dependence of the
synchronization time $\tau_{\rm sync}$ on the network topology nor on the number of large ($N_P$) and small ($N_R$) power generators. The mean
value of $\tau_{\rm sync}$ is always slightly larger than the relaxation constant $\alpha^{-1}$. Furthermore, also the standard deviation
of $\tau_{\rm sync}$ for different realizations of the random networks is only maximum 3 percent of the mean value. A significant influence of the network structure on the
synchronization time has been found only in the weak damping limit, i.e. for very large values of $P_0/\alpha$ and $K/\alpha$.

\begin{figure}[t]
\centering
\includegraphics[width=8.4cm, angle=0]{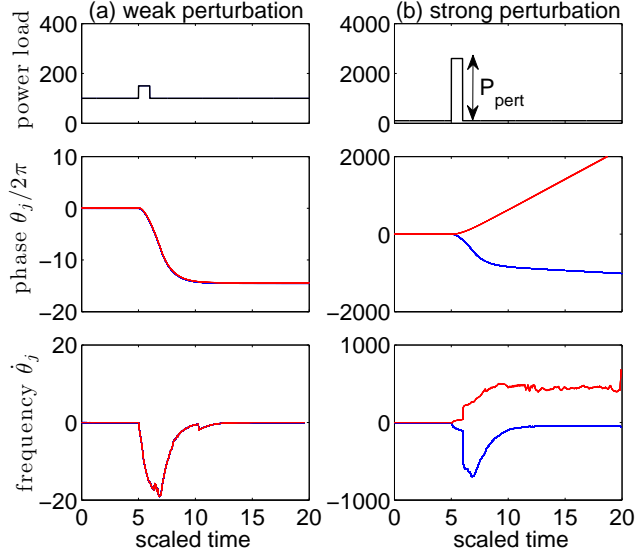}
\caption{\label{fig:robustness1}
Weak and strong perturbation. The upper panels show the time-dependent power load of the consumers. A perturbation of strength $P_{\rm pert}$ is
applied in the time interval $t \in [5,6]$. The lower panels show the resulting dynamics of the phase $\theta_j$ and the frequency $\dot \theta_j$
of the consumers (blue lines) and the power plants (red lines). The dynamics relaxes back to a steady state after the perturbation for a weak
perturbation (a), but not for a strong perturbation (b). In both cases we assume a regular grid with $N_P = 10$.
}
\end{figure}

\subsection{Stability against perturbations}

\begin{figure}[t]
\centering
\includegraphics[width=10cm, angle=0]{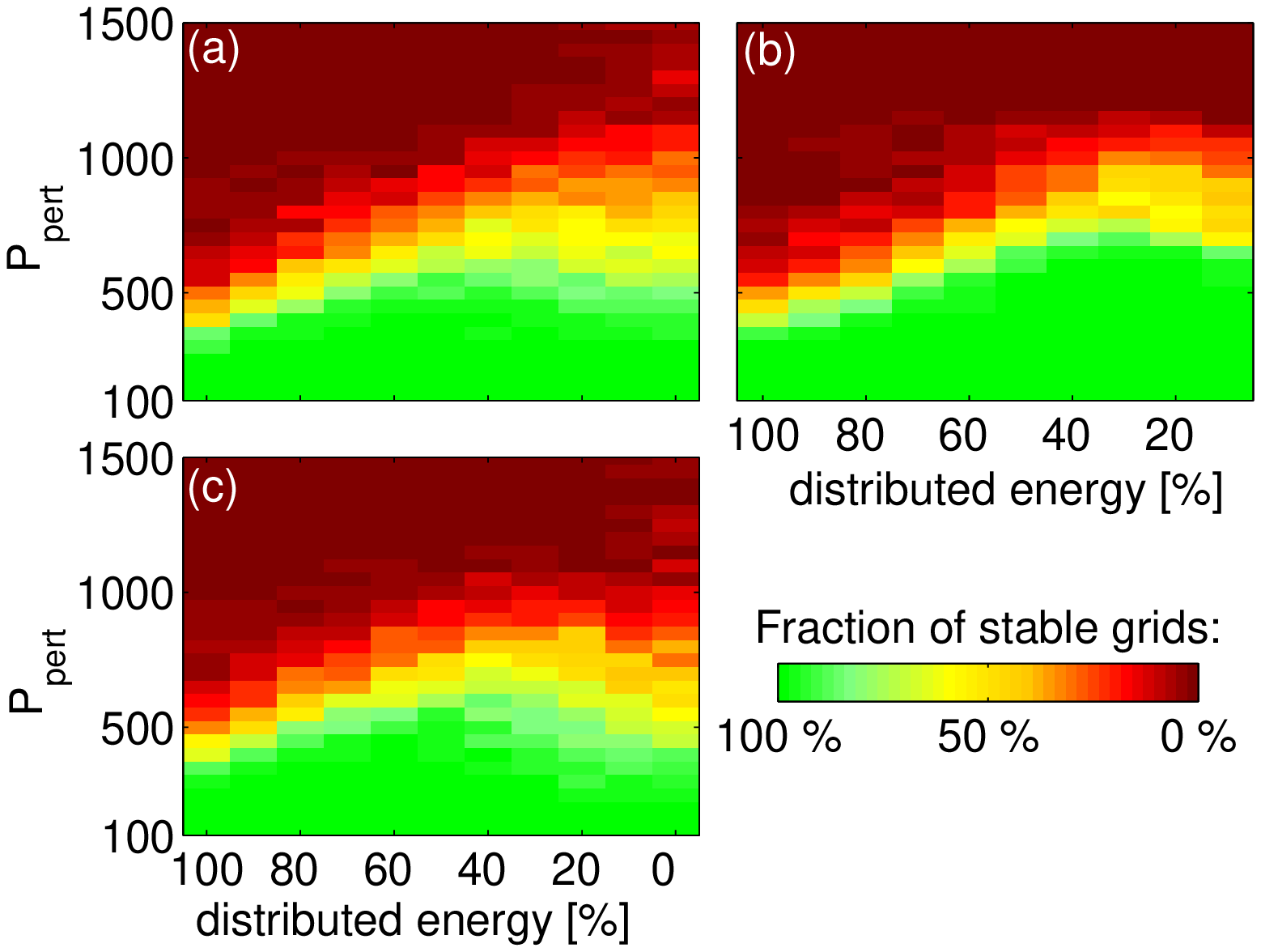}
\caption{\label{fig:robustness2} 
Robustness of a power grid. The panels show the fraction of random grids which are unstable against a perturbation as a function of the
perturbation strength $P_{\rm pert}$  and the fraction of decentralized energy $1 - N_P/10$. (a) Quasi-regular grid, (b) random network and
(c) small-world network.
}
\end{figure}

Finally, we test the stability of different network structures against perturbations on the consumers side. We perturb the system after
it has reached a stable state and measure if the system relaxes to a steady state after the perturbation has been switched off
again. The perturbation is realized by an increased power demand of each consumer during a short time interval ($\Delta t = 10s$) as
illustrated in the upper panels of Fig.~\ref{fig:robustness1}. Therefore the condition of \ref{eqn:sum} is violated and the system cannot remain in its stable state. After the perturbation is switched off again, the system relaxes
back to a steady state or not, depending on the strength of the perturbation. Fig.~\ref{fig:robustness1} shows examples of the dynamics for a weak
(a) and strong (b) perturbation, respectively.

These simulations are repeated 100 times for every value of the perturbation strength for each of the three network topologies. We then count the fraction of
networks which are unstable, i.e do not relax back to a steady state. The results are summarized in Fig.~\ref{fig:robustness2} for different network
topologies. The figure shows the fraction of unstable grids as a function of the perturbation strength and the number of large power plants. For all
topologies, the best situation is found when the power is generated by both large power plants and small power generators. An explanation is that
the moment of inertia of a power source is larger if it delivers more power, which makes it more stable against perturbations. On the other hand,
a more distributed arrangement of power stations favors a stable synchronous operation as shown in Sec. \ref{sec-starnet}.

Furthermore, the variability of the power grids is stronger for low values of $N_P$, i.e. few large power plants. The results do not change much
for networks which many power sources (i.e. high $N_P$) because more power sources are distributed in the grid. Thus the random
networks differ only weakly and one observes a sharp transition between stable and unstable. This is different if only few large power plant are
present in the network. For certain arrangements of power stations the system can reach a steady state even for strong
perturbations. But the system can also fail to do so with only small perturbations if the power stations are clustered. This emphasizes the
necessity for a careful planning of the structure of a power grid to guarantee maximum stability.

\section{Conclusion and Outlook}

In the present article we have analyzed a dynamical network model for the dynamics of a power grid. Each element of the network is modeled as a second-order
oscillator similar to a synchronous generator or motor. Such a model bridges the gap between a microscopic description of electric machines and
static models of large supply networks. It incorporates the basic dynamical effects of coupled electric machines, but it is still simple enough to
simulate and understand the collective phenomena in complex network topologies.

The basic dynamical mechanisms were explored for elementary network structures. We showed that a self-organized phase-locking of all generators and
motors in the network is possible. However, this requires a strong enough coupling between elements. If the coupling is decreased, the synchronized
steady state of the system vanishes.

We devoted the second part to a numerical investigation of the dynamics of large networks of coupled generators and consumers, with an
emphasis on self-organized phase-locking and the stability of the synchronized state for different topologies. It was shown that the critical
coupling strength for the onset of synchronization depends strongly on the degree of decentralization. Many small generators can
synchronize with a lower coupling strength than few large power plants for all considered topologies. The relaxation time to the steady state, however, depends only weakly on
the network structure and is generally determined by the dissipation rate of the generators and motors. Furthermore we investigated the robustness
of the synchronized steady state against a short perturbation of the power consumption. We found that networks powered by a mixture of small
generators and large power plants are most robust. However, synchrony was lost only for perturbations at least five times their normal energy consumption
in all topologies for the given parameter values.

For the future it would be desirable to gain more insight into the stability of power grids regarding transmission line failures, which is not fully
understood yet \cite{Kurths13}. For instance, an enormous challenge for the construction of future power grids is that wind energy sources are
planned predominantly at seasides such that energy is often generated far away from most consumers. That means that a lot of new transmission lines wil be added
into the grid and such many more potential transmission line failures can occur. Although the general topology of these future power grids seem to be not that
decisive for their functionality, the impact of including or deleting single links is still not fully understood and unexpected behaviors can occur \cite{Witthaut}. Furthermore it is highly
desirable to gain more inside into collective phenomenma such as cascading failures to prevent major outages in the future.


\end{document}